\documentclass[twocolumn,trackchanges]
{aastex631}

\usepackage{float}
\usepackage{chngpage}

\usepackage{tikz}
\usetikzlibrary{arrows,calc,positioning}

\tikzstyle{intt}=[draw,text centered,minimum size=6em,text width=5.25cm,text height=0.34cm]
\tikzstyle{intl}=[draw,text centered,minimum size=2em,text width=2.75cm,text height=0.34cm]
\tikzstyle{int}=[draw,minimum size=2.5em,text centered,text width=3.5cm]
\tikzstyle{intg}=[draw,minimum size=3em,text centered,text width=6.cm]
\tikzstyle{sum}=[draw,shape=circle,inner sep=2pt,text centered,node distance=3.5cm]
\tikzstyle{summ}=[drawshape=circle,inner sep=4pt,text centered,node distance=3.cm]
\usetikzlibrary{shapes.geometric, arrows}
\tikzstyle{arrow} = [thick,->,>=stealth]

\usepackage{enumitem}

\usepackage{orcidlink}
\usepackage{multirow}
\shorttitle{Anomalous Transits}
\shortauthors{Zuckerman et al.}

\begin{document}

\title{The Breakthrough Listen Search for Intelligent Life: Detection and Characterization of Anomalous Transits in Kepler Lightcurves}

\correspondingauthor{Zuckerman}
\email{anna\_zuckerman@alumni.brown.edu}

\newcommand{\UCB}{Breakthrough Listen,  University of California Berkeley, Berkeley CA 94720}
\newcommand{\UCBAstroandGeo}{Departments of Astronomy and Earth and Planetary Science, University of California Berkeley, Berkeley CA 94720}
\newcommand{\seti}{SETI Institute, Mountain View, California}
\newcommand{\UW}{Department of Astronomy, University of Washington, Seattle WA, 98195}
\newcommand{\Brown}{Department of Physics, Brown University, Box 1843, 182 Hope St., Providence, RI 02912}
\newcommand{\CUB}{Department of Astrophysical and Planetary Sciences, University of Colorado Boulder, 2000 Colorado Ave, Boulder, CO 80309}
\newcommand{\FOOTBALLERS}{Department of Physics and Astronomy, University of Manchester, UK}
\newcommand{\KZA}{University of Malta, Institute of Space Sciences and Astronomy, Msida, MSD2080, Malta}
\newcommand{\NTargets}{228}
\newcommand{\NSearched}{218}
\newcommand{\NSearchedTransits}{39879}
\newcommand{\NMissing}{6}
\newcommand{\NDeep}{92}
\newcommand{\NTTV}{51}

\author[0000-0002-2412-517X]{Anna Zuckerman} 
\affiliation{\UCB}
\affiliation{\CUB}

\author[0000-0002-0637-835X]{James R. A. Davenport}
\affiliation{Astronomy Department, University of Washington, Box 951580, Seattle, WA 98195, USA}

\author[0000-0003-4823-129X]{Steve Croft}
\affiliation{\UCB}
\affiliation{\seti}

\author[0000-0003-2828-7720]{Andrew Siemion}
\affiliation{\UCB}
\affiliation{\seti}
\affiliation{\FOOTBALLERS}
\affiliation{\KZA}

\author[0000-0002-4278-3168]{Imke de Pater}
\affiliation{\UCBAstroandGeo}




\begin{abstract}
Never before has the detection and characterization of exoplanets via transit photometry been as promising and feasible as it is now, due to the increasing breadth and sensitivity of time domain optical surveys. Past works have made use of phase-folded stellar lightcurves in order to study the properties of exoplanet transits, because this provides the highest signal that a transit is present at a given period and ephemeris. Characterizing transits on an individual, rather than phase-folded, basis is much more challenging due to the often low signal-to-noise ratio (SNR) of lightcurves, missing data, and low sampling rates. However, by phase-folding a lightcurve we implicitly assume that all transits have the same expected properties, and lose all information about the nature and variability of the transits. We miss the natural variability in transit shapes, or even the deliberate or inadvertent modification of transit signals by an extraterrestrial civilization (for example, via laser emission or orbiting megastructures). In this work, we develop an algorithm to search stellar lightcurves for individual anomalous (in timing or depth) transits, and we report the results of that search for \NSearched\ confirmed transiting exoplanet systems from Kepler.

\end{abstract}



\section{Introduction} \label{sec:intro}

The expanding availability of photometric stellar lightcurves, for instance from the Kepler and Transiting Exoplanet Survey Satellite (TESS) missions, has provided the chance to discover and understand the orbital properties of thousands of exoplanet systems in recent years.
Past studies have been limited to detection of transiting exoplanets or characterization of planetary and orbital properties from analysis of collectively phase-folded transits. The phase-folded shape of an exoplanetary transit provides a wealth of information about the planetary system that produced the transit, but can inherently only provide insight into the ensemble properties of all the observed transits of that planet. While some work has been dedicated to understanding the provenance of unusually shaped transits by reverse-engineering the shape of a transiting body from the transit \citep{Sandford_2019}, this has always been done after phase-folding. No past work has comprehensively searched for and characterized missing or anomalous individual transits in stellar lightcurve data. 

The population variability of exoplanet transits and the nature and prevalence of unexpected transits is poorly understood. This represents a major gap in the study of exoplanet populations. For instance, individual transits can be a powerful tool in studying stellar surface variability such as star spots \citep{Morris2017}. In addition, unexpected or anomalous transits represent an important potential technosignature in the Search for Extraterrestrial Intelligence (SETI) \citep{Wright2019}. Intelligent extraterrestrial civilizations may  broadcast their presence by intentionally or inadvertently modifying the shape of their host planet's transit, for example with laser emission or megastructures \citep[e.g.][]{kipping2016,arnold2005}. In addition, observations discrepant with expectation often drive theoretical advancement, so the discovery of unusually-shaped transits produced by unexplained phenomena would be scientifically interesting in and of itself and would necessitate new concepts in planetary or stellar astrophysics, if not SETI.

We present a new algorithm to analyze lightcurves on a transit-by-transit basis without phase-folding, in order to study the variability between individual transits and search for transits that are missing or anomalous. We present the methodology and results of a search for transits which are apparently missing, or which exhibit unexpectedly large transit-timing-variations (TTV's) or apparent depths. We find no transits with anomalous properties which cannot be explained after visual vetting.

\section{Data, Pre-processing, and Target Selection} 
\label{sec:data}

\subsection{Data}
\label{subsec:data}

The Kepler mission \citep{Borucki2010} has produced 2708 confirmed detections of transiting exoplanet systems\footnote{As of June 6, 2022, as reported by the NASA Exoplanet Archive (https://exoplanetarchive.ipac.caltech.edu)}. The program ran from 2009 to 2013, observing approximately 150,000 stars in multiple 90-day quarters, at a cadence of either 30 minutes or 60 seconds. The mission prioritized Main Sequence stars for which Earth-like planets would be detectable \citep{Batalha_2010}. Lightcurves produced by the Kepler Mission Science Operations Center can be publicly downloaded from the Barbara A. Mikulski Archive for Space Telescopes (MAST) archive (\dataset[DOI: 10.17909/T9059R]{https://doi.org/10.17909/T9059R}). 
The Kepler Science Processing Pipeline is described in \cite{Jenkins2010}. Lightcurves comprise measured flux as a function of time. They often contain extended intervals of missing data (due to the telescope entering safe mode, rotating towards Earth, or executing a quarterly roll) as well as individual data points flagged for quality issues (due to cosmic ray hits, reaction wheel zero crossings, impulsive outliers, thruster firings, etc.). 

\subsection{Pre-processing}
\label{subsec:pre-processing}

\begin{figure*}[ht]
\centering
\includegraphics[width=\textwidth]{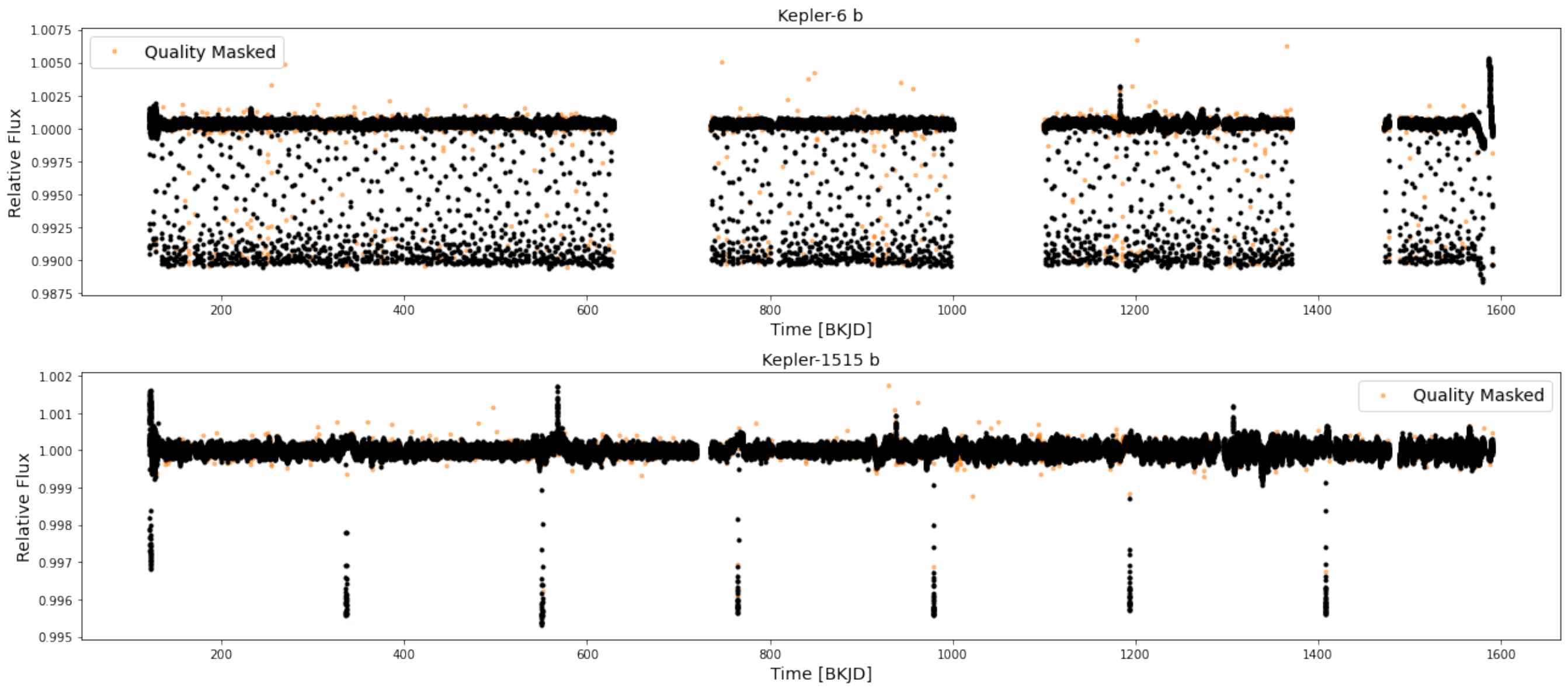}
\caption{Example detrended, stitched, and masked lightcurves. Orange points are data that have been removed from our analysis due to quality issues flagged by the Kepler pipeline.}
\label{fig:stitch}
\end{figure*}

Before transits can be identified and anomalies detected, we first process each target lightcurve. For each planetary system, we download lightcurve data from the MAST using the interface provided by the \texttt{LightKurve} package. We use the default Presearch Data Conditioning Simple Aperture Photometry (PDCSAP) flux values that have been cotrended by the Kepler team using the PDC cotrending algorithm \citep{Stumpe2012, Smith2012, Stumpe2014}. Kepler's observing run is divided into quarters, punctuated by rolls of the telescope. We fit and remove a linear trend from each quarter, and stitch the quarters into one continuous lightcurve (see Figure \ref{fig:stitch}). We then mask out data with quality issues flagged during Kepler data acquisition. Each lightcurve corresponds to one stellar target which may or may not be host to more than one exoplanet. So, for each target planet we mask out any additional transits from ``sibling planets'', which might interfere with our fitting process. This masking uses a folding and Box-Least-Squares fitting approach, and so relies on the first order assumption that all transits appear near their expected locations. We acquire a list of sibling planets by querying the NASA Exoplanet Archive database (\dataset[DOI 10.26133/NEA4]{https://catcopy.ipac.caltech.edu/dois/doi.php?id=10.26133/NEA4})\footnote{Accessed on 2022-01-22 at 22:20 PST, returning 2363 rows}.
On very rare occasions, key parameters returned from the NASA exoplanet archive were null, preventing the masking of a sibling planet. In these cases, sibling transits that are not properly masked can interfere with the fitting. This failure mode is vetted in Section \ref{sec:results} by querying the Simbad archive when additional siblings are suspected.

\subsection{Target Selection}
\label{subsec:target_selection}

Because we analyze transits individually, we are strongly limited by the depth and duration of the transit signal relative to the local noise. In order to maximize the statistical significance of any characterization we make, we sort all confirmed Kepler targets by the ratio of the expected transit depth to the scatter in the flux (defined as the the median rolling standard deviation (SD) of the flux values, with a window size of 10 timesteps), and take only the best 10\% of targets. This corresponds to a threshold ratio of 6.73. We choose 10\% as our threshold because targets below this cutoff tend to have transits with depths similar to the lightcurve noise (less than a few times the noise), and thus are not good candidates for unambiguous detection of missing transits. If the expected depth for a particular transiting system is near or below the average scatter of the lightcurve, we cannot make any significant statement about the nature of an individual transit. This produces \NTargets\ targets. We determine the expected depth by fitting the ensemble of transits together for each planet to produce an average transit model for that planet. We download the reported transit properties (period, semimajor axis, inclination, eccentricity, argument of periastron, impact parameter, transit duration, ephemeris, and planet and star radii) for each planetary system from the NASA Exoplanet Archive, and use these as initial parameters for fitting. We perform the transit fitting using a \texttt{Batman} transit model wrapped in Python's \texttt{lmfit} method. The \texttt{Batman} package (described fully by \citealt{Kreidberg_2015}) produces a transit model based on input planetary and orbital parameters, and \texttt{lmfit} allows us to fit for the best \texttt{Batman} transit model using a brute-force method over a grid of points. The ensemble fit parameters sometimes differ significantly from the reported NASA transit parameters. Because we are concerned only with deviations in the transit depth or midpoint for individual transits, we are not concerned with defining physically accurate transit parameters, but rather with defining the model that best represents the depth and timing of transits for each planetary system as a whole. We take the depth of this ensemble model as the expected transit depth for each individual transit. By selecting the systems with the highest ratio of expected transit depth to flux scatter, we aim to produce the highest confidence results possible.

\section{Methods} \label{sec:methods}
\subsection{Expected Transit Parameters} 
\label{subsec:expected_params}

Before fitting individual transits, we must first define a set of expected parameters to serve as initial guesses for individual transit fits. We take the preliminary set of parameters determined for the purpose of target selection (as described in Section \ref{subsec:target_selection}), and inspect the fits produced by those parameters by eye. In the case of a poor fit, we manually update the fit parameters to produce a better expected transit shape. This allows us to have high confidence that the expected transit models well represent an `average' transit, and thus provide a good baseline from which to define unusual transit shapes. We refer to this set of parameters as the `ensemble fit parameters', because they provide our best fit to the entire ensemble of transits for a planet.

\subsection{Anomaly Search}
\label{subsec:fitting}

After pre-processing, the lightcurves are ready to be searched for missing or unusual transits. For each star, we iterate over each transit. For each expected transit midpoint (with midpoints determined using the period and ephemeris provided by the NASA Exoplanet Archive in the Kepler KOI table \citep{Thompson2018}, we define a window spanning three times the expected transit duration on either side of the expected midpoint. If insufficient data (defined as less than 70\% of the expected number of datapoints in the window) is present in a given transit's window due to data gaps or data quality issues, we ignore that transit. Even so, missing or sparse data is one of the most common reasons for poor fitting of individual transits.

We then fit the transit with a \texttt{Batman} transit model. We first mask the expected location of the transit in order to fit and remove a linear trend as a first-order correction for stellar variability. Short-timescale non-linear stellar variability is not removed, and this is another common reason for poor fitting. Next, we produce a preliminary model of the transit in order to update the expected midpoint, and repeat the process of masking the transit and removing a linear trend. This helps avoid errors in baseline trend removal when the transit is not centered at the expected location. We then fit a final \texttt{Batman} model to the transit to produce a set of fit parameters, using the ensemble model parameters as initial guesses. 

As a reference of comparison for individual transits, we take the median of each parameter produced in this final \texttt{Batman} fit over all transits and refer to this set of parameters as the `median fit parameters'. This set of median fit parameters provides the standard against which we define which individual transits are considered anomalous. 

Once each transit has been fit, we flag the transits that meet certain requirements. We define three flags: (1) transits which appear ``missing", (2) transits which appear unusually deep, and (3) transits which appear to have unexpectedly large transit-timing variations (TTVs). These are defined quantitatively as follows: 
\begin{enumerate}[label={(\arabic*)}]
  \item fit transit depth shallower than 3 times the flux SD, and greater than 3 sigma away from the median of the fit transit depths (the transit is effectively indistinguishable from a flat section of lightcurve).
  \item fit transit depth exceeds the median fit transit depth by more than 3 sigma.
  \item fit transit TTV exceeds the median fit TTV by more than 5 sigma. We choose 5 sigma in this case to prevent flagging expected large TTVs due to coherent variation in transit timing from sibling planets.
\end{enumerate}

\begin{figure*}[ht]
\centering
\includegraphics[width=0.7\textwidth]{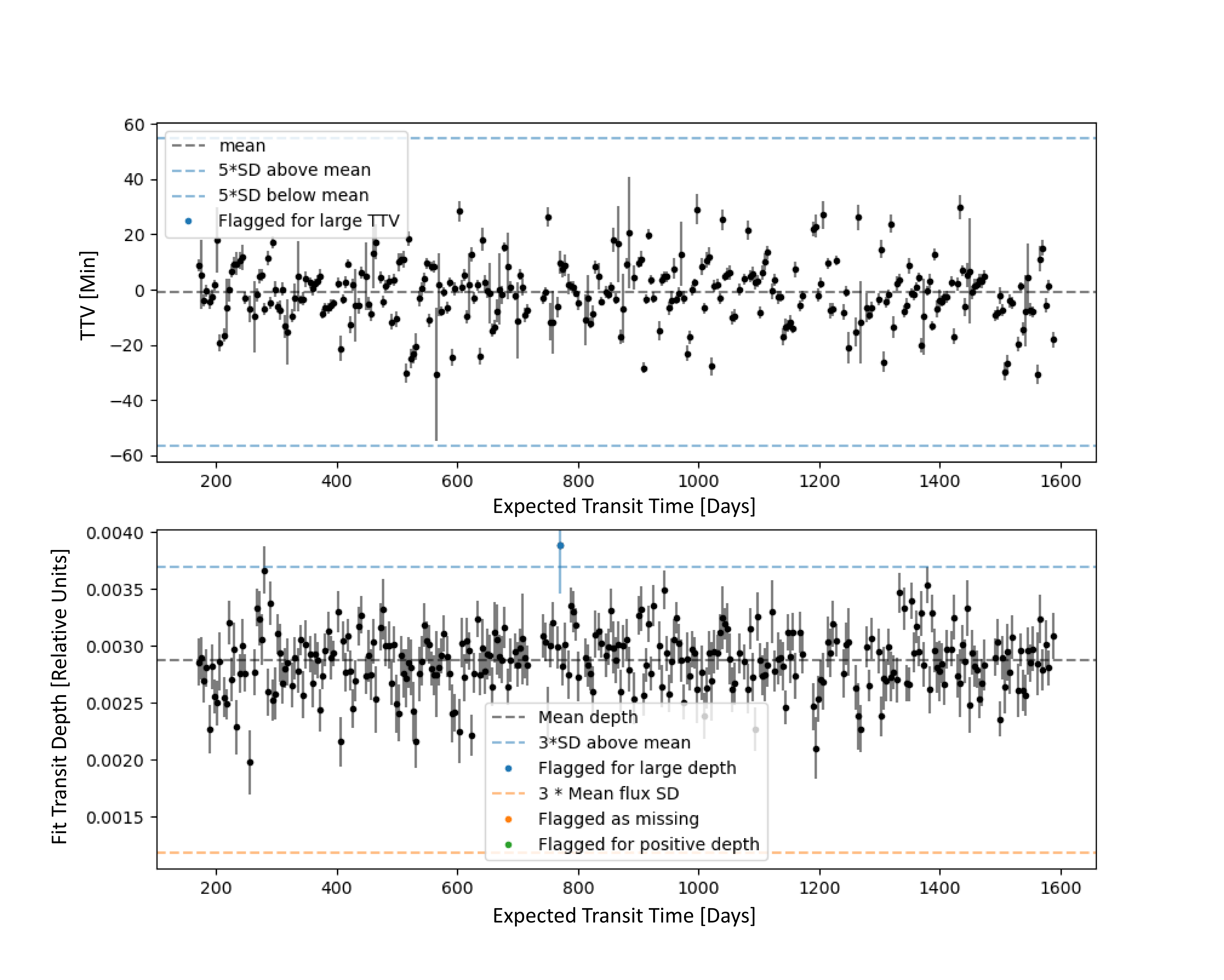}
\caption{Demonstration of the thresholds used for flagging fit transit parameters for an example planet (Kepler-696b). Each datapoint represents one transit.}
\label{fig:thresholds}
\end{figure*}

Defining the flagging criteria based on the actual distribution of fitted transit properties circumvents issues with small discrepancies between the ensemble and mean fit parameters which sometimes arise due to imperfect expected parameter values. It also allows us to flag TTVs even in several-planet systems which are often expected to have large TTVs due to the gravitational effects of planets on their siblings' orbits. Figure \ref{fig:thresholds} demonstrates the threshold applied for each flag.

\subsection{Signal Injection and Recovery}
 
 In order to assess our confidence in flags produced by our anomaly search for any given planetary system, we perform an injection and recovery of simulated transits and evaluate both the accuracy with which we can recover their depths and timings, and the accuracy of the flagging processes. 
 We would like to be able to reject flagged transits in systems for which the injection and recovery shows it is difficult to accurately determine the transit parameters, or systems for which many transits are erroneously flagged. 
 
 We inject transits into each lightcurve as follows. First, we define a set of 1000 injection locations chosen randomly across the lightcurve under the requirement that sufficient data is present in the surrounding window to perform a \texttt{Batman} fit. We then choose a sample randomly from a distribution of depths and TTVs without replacement to produce the injection parameters. The depth parameter space spans from zero to 3.5 sigma above the median fit depth, and the TTV parameter space spans from zero to 1.1 times the size of the fitting window. We choose this range to ensure that we sample from all regions which should produce flags, as well as regions expected to erroneously produce flags. We inject each transit by defining a \texttt{Batman} model using the chosen depth and TTV, and median fit parameter values for the other parameters, and add the injection model onto the existing lightcurve in order to preserve noise and stellar variability. 
 
\begin{figure*}[ht]
\centering
\includegraphics[width=0.85\textwidth]{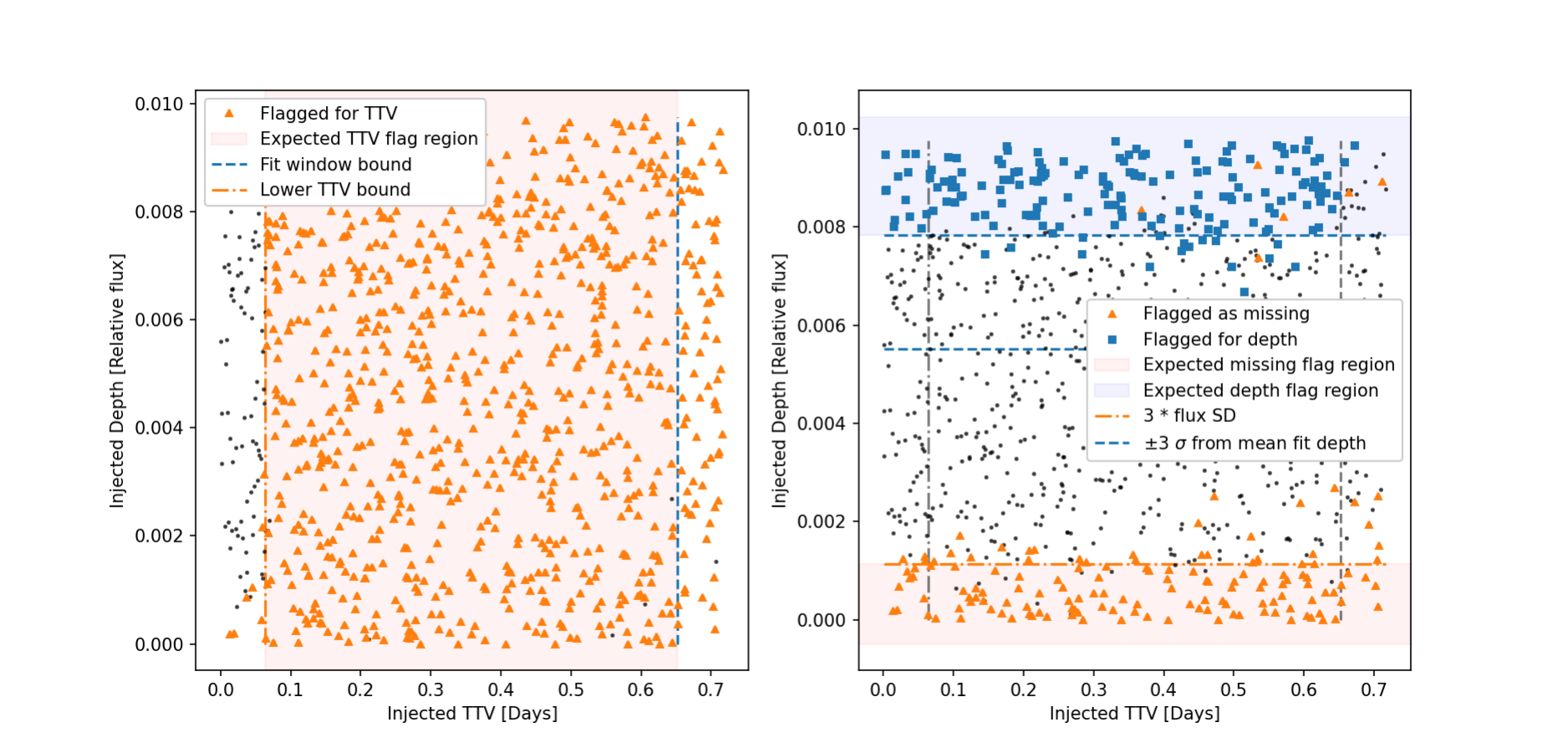}
\caption{Scatter plot demonstrating injection and recovery accuracy for a sample lightcurve. Each point represents the injected properties of recovered transits. In the left-hand plot, orange triangles represent transits flagged for large TTVs, and the orange filled area represents the region of parameter space which should produce a TTV flag. In the right-hand plot, orange triangles represent transits flagged as missing, blue squares represent transits flagged for large depths, and the orange and blue regions represent the corresponding regions of parameter space expected to produce flags. Not all transits injected in the shaded regions are flagged due to imperfect recovery of the injected properties.}
\label{fig:inj_rec}
\end{figure*}

 We then attempt to recover each injection by performing the fitting exactly as described in Section \ref{subsec:fitting}. Once all injections are fit the flags are defined exactly as described above, using the median and standard deviation of the real individual transit fits to determine the threshold values for flagging. We calculate the RMS error between the injected and recovered depths and TTVs for each planet, and produce diagnostic plots to visualize the recovery accuracy (as shown in Figure~\ref{fig:inj_rec}). For this paper, we define `true positive' injections to be injections whose injected depth or TTV is within the region that should produce a flag, and whose recovered depth or TTV does in fact produce the expected flag. We define `true negative' flags analogously, as transits whose injected parameters should not produce a flag, and whose recovered parameters indeed are not flagged. `False positive' injections are those whose parameters should not produce a flag, yet the recovery is inaccurate and the recovered transit is flagged. Similarly, `false negative' injections are those whose parameters should produce a flag, but due to poor recovery are not flagged. 
 
 We perform this injection and recovery process for each planetary system. We can then use the accuracy of the injection recovery to inform vetting of flagged transits.

\section{Results}
\label{sec:results}

We run the individual transit fitting algorithm described in Section \ref{subsec:fitting} on each of the \NTargets\ planetary targets described in Section \ref{sec:data}. The \texttt{Batman} model fails to converge for 10 of these targets, leaving \NSearched\ which we can search for anomalous transits. This means we search \NSearchedTransits\ total transits. Using the flags defined above, we flag \NMissing\ transits as apparently missing, \NDeep\ as unexpectedly deep, and \NTTV\ as having unexpectedly large TTVs. 

We then visually inspect these flagged transits. We reject flagged transits using several criteria, which vary slightly for each flag. In fitting individual transits, we are often challenged by the sparsity of data around the expected transit. Thus, all flags have some false positives when the transit is truly present in the data with its expected shape, but a good fit to the transit cannot be produced due to missing or sparse data. Another complication for all flags arises from sibling transits which fall within the fitting window. Though sibling transits are masked before fitting, in a few cases sibling transits with large TTVs cannot be perfectly masked beforehand and can interfere with fitting. 

\subsection{Transits Flagged as Missing}
\label{subsec:{missing}} 
Our primary interest is to identify and examine potentially ``missing" transits. In general, we find that despite the noisiness and sparseness of data for individual transits, we are able to accurately determine that a transit is present at each expected location and to fit its depth and TTV. However, some transits are flagged as missing.

\begin{figure*}[ht]
\centering
\includegraphics[width=\textwidth]{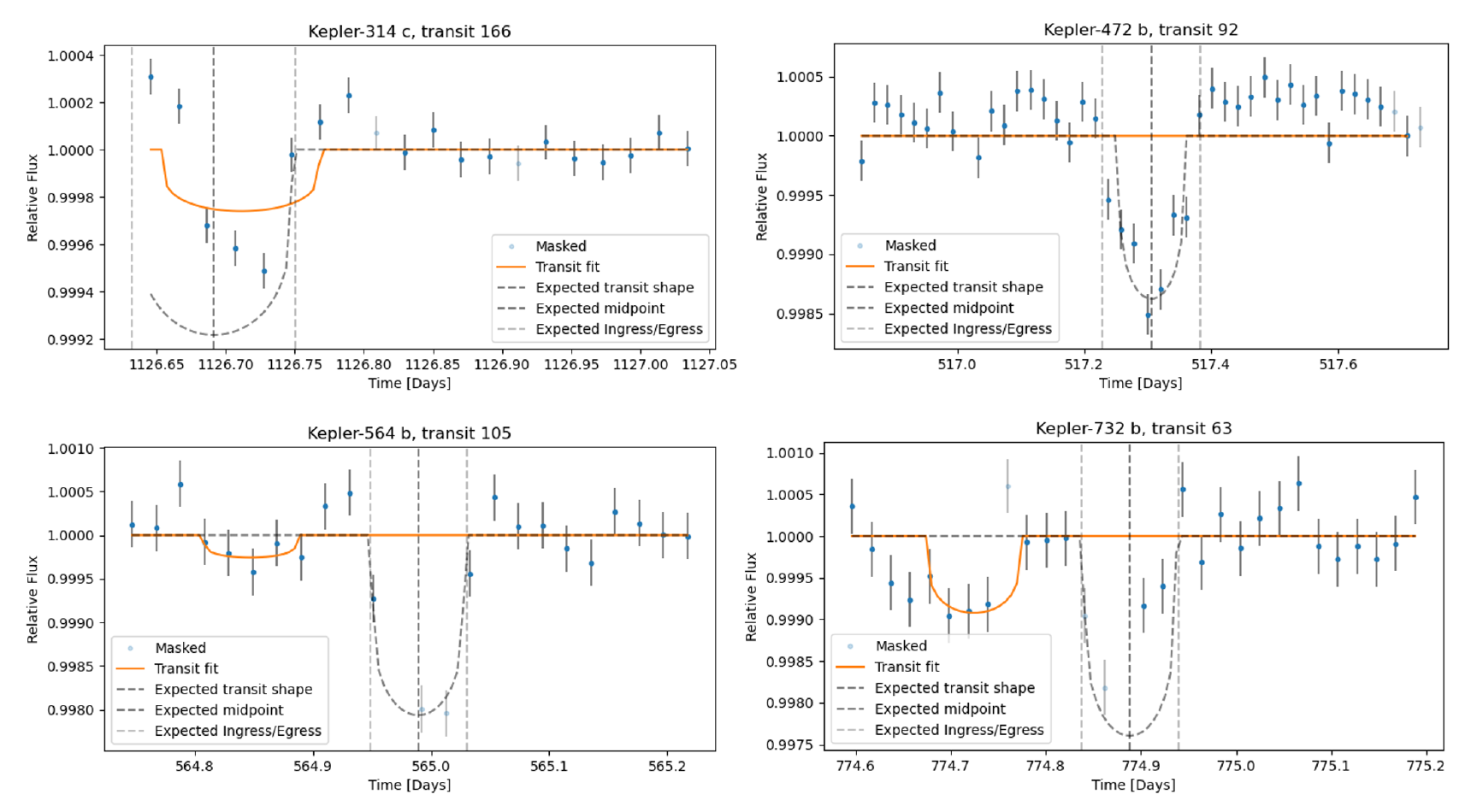}
\caption{Four of the six transits flagged as missing are clearly present on visual inspection of the data. Top left: Kepler-314c, transit 166, is flagged as missing but is clearly visible in the raw data. Missing data at the edge of the fit window and short timescale stellar variability make the fitting difficult and produce a false positive. Top right: Kepler-472, transit 92 is flagged as missing yet is clearly present on visual inspection. Bottom left: Kepler-564, transit 105 is mis-fit due to two quality masked data points exactly at the transit minimum, which effectively hide the transit. Bottom right: Similarly, the model fits adjacent stellar variability instead of transit 63 of Kepler-732b because of a single missing datapoint in the fit, as well as short-timescale stellar variability which alters the shape of the transit.}
\label{fig:missing_bad_data}
\end{figure*}

For such missing transits, one possible false positive comes from poor fitting due to missing or sparse data in the region around the transit. Four of the six transits flagged as missing are clearly present in the data on visual inspection, but are fit poorly due to missing data and/or stellar variability. These are shown in Figure~\ref{fig:missing_bad_data}.

\begin{figure*}[ht]
\centering
\includegraphics[width=\textwidth]{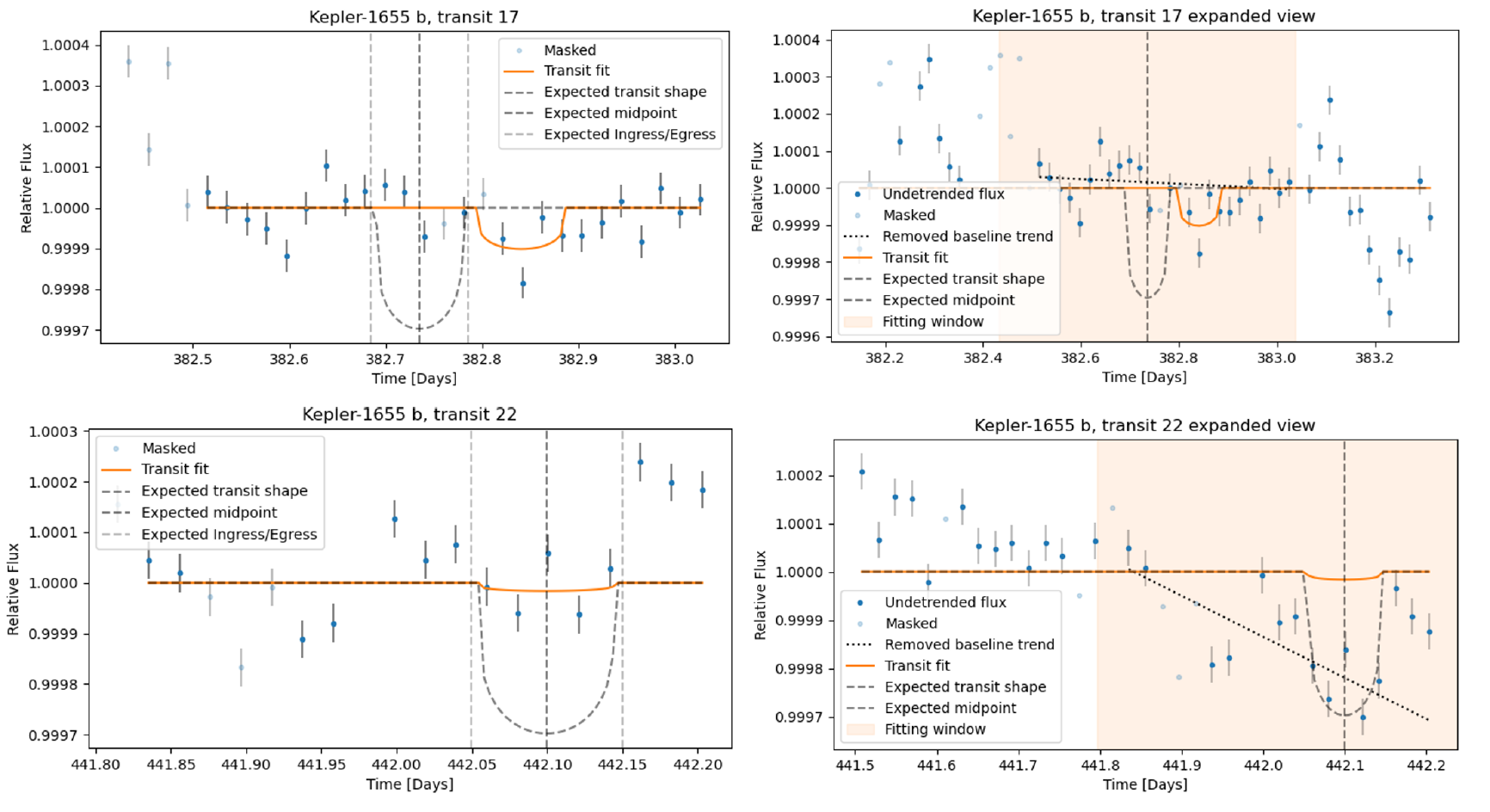}
\caption{Kepler-1655b transits 17 (top) and 22 (bottom) are flagged as "missing" and are not obviously present on visual inspection, but are ultimately rejected due to poor linear trend removal and due to stellar variability which could mask a present transit. }
\label{fig:missing}
\end{figure*}

The two remaining candidates are both transits of the planet Kepler-1655b (Figure~\ref{fig:missing}). For these two, the transit is not visually identifiable in the processed data. The first step for further visual vetting is an inspection of the unprocessed (i.e., with no linear detrending) lightcurve in a wider window around the flagged transit. When inspected this way, transit 22 has a visually identifiable feature at the location of the fit transit which matches the expected transit shape, and this can be identified as the transit itself. A wide section of missing data beginning three datapoints to the right of the transit egress, in conjunction with local stellar variability, impairs accurate linear trend removal. The trend that is removed artificially raises, and therefore obscures, the transit.
For transit 17, the cause of the flag is somewhat less clear. However, inspecting the wider region around the transit reveals stellar variability on the same order as the expected transit depth. This is an edge case for which the timescale of the stellar variability was too short to be detrended, yet too long to be included in the calculation of the SD. Thus we must also reject this transit as a statistically significant``missing" transit.

Though we are most interested in potentially missing transits, our dataset is also highly amendable to a search for transits with unusually large depths or TTVs. These are discussed in the following two sections.
 
\subsection{Transits Flagged as Unexpectedly Deep}
\label{subsec:{deep}} 

One possible false positive for transits flagged as unexpectedly deep comes from poor fitting due to missing or sparse data. In the case of missing data, there is an additional failure mode that occurs when the midpoint is fit inside a region of missing data, and the depth is then essentially unbounded. Of the \NDeep\ flagged transits, 23 are rejected as false positives for this reason. An additional 31 are rejected because non-linear stellar variability results in a mischaracterization of the local baseline, and thus an apparently deeper transit. Also, 19 transits are rejected because the scatter in the out-of-transit lightcurve is large compared to the difference between the fit depth and expected depth; thus the excess depth could be explained by random variability rather than a true increase in transit depth. Large instrumental scatter can also make the linear detrending unreliable for these transits.

Another class of false positive comes from sibling transits which overlap with the target transit, deepening the apparent transit. Though we attempt to mask these sibling transits, we are unable to initially mask a sibling transit properly in cases when the transit exhibits a very large TTV, or in the rare case that a search of the NASA exoplanet archive returns null parameters. We vet for the first failure mode by overplotting the individually fit midpoints of resolved sibling planets, and reject a single transit (Kepler-247c transit 113) which clearly has an overlapping sibling. To check for the second failure mode, we manually search the Simbad database for siblings and in two cases find an unmasked sibling planet. From this we can reject five flagged transits of Kepler-18c, because visual inspection reveals the transits of Kepler-18d clearly move in and out of the fitting window. Similarly, we reject Kepler-51b transit 30 because an additional sibling (Kepler-51d) is reported in Simbad and has an expected transit midpoint that directly overlaps. 

We reject another four transits because analysis of the injection and recovery process reveals that accurate recovery of transit parameters is difficult. If for a given system we are unable to recover injected deep transits, or we flag normal transits as unusually deep, we cannot have high confidence in any transit flagged for that system and so must reject it. We impose a threshold on the fraction of false positive injections, out of the total number of injections which should not be flagged. We choose that threshold to be 10\% in order to reject the clear tail of outliers with especially poor injection recovery while retaining the majority of transits. This test allows us to reject transits of Kepler-25c, Kepler-17b, Kepler-685b, and Kepler-506b. This leaves nine transits. 

The next step for visual vetting is an inspection of the unprocessed (i.e., with no linear detrending) lightcurve flux in a wider window around the flagged transit. When inspected this way, several transits can be clearly rejected based on one of the previously identified failure modes. When the pre-detrending lightcurve is inspected in a wider window, it becomes apparent that Kepler-314c transit 109 and Kepler-682b transit 41 have short-timescale stellar variability with a timescale comparable to the transit duration and with amplitude comparable to the difference between the fit and expected depth. Thus, the apparently large depth could be caused by unfortunately timed stellar variability.

This leaves five remaining transits (Kepler-685b transit 791, Kepler-422b transit 99, Kepler-74b transit 56, Kepler-433b transit 248, and Kepler-41b transit 616). Although these transits do not fall under any of the failure modes described above, they are not anomalous enough to warrant further investigation. All five of these transits appear to fall on the tail of a normal distribution of fit depths for their respective systems, or result from small coherent trends in an otherwise normal fit depth distribution (which may occur due to moving starspots or other stellar surface features).

\subsection{Transits Flagged for Large TTVs}
\label{subsec:{ttv}} 

As in the previous two cases, for transits flagged for large TTVs a possible false positive comes from poor fitting due to missing or sparse data near the transit. We reject 27 of the \NTTV\ transits flagged for large TTVs for this reason. We reject another 12 that were fit poorly due to short timescale stellar variability or single missing datapoints that bias the fitting. For a further set of five, short timescale stellar variability produces a feature that mimics the expected transit shape enough to `fool' the \texttt{Batman} model into fitting the variability rather than the transit itself.

Another class of false positive comes from sibling transits present in the fitting window which are fit instead of the target transit. We reject the same five transits of Kepler-18c described in Section~\ref{subsec:{deep}} because a Simbad search reveals a sibling planet unresolved in the NASA Exoplanet Archive whose transits coincide with the flagged transits.

\begin{figure*}[ht]
\centering
\includegraphics[width=0.5\textwidth]{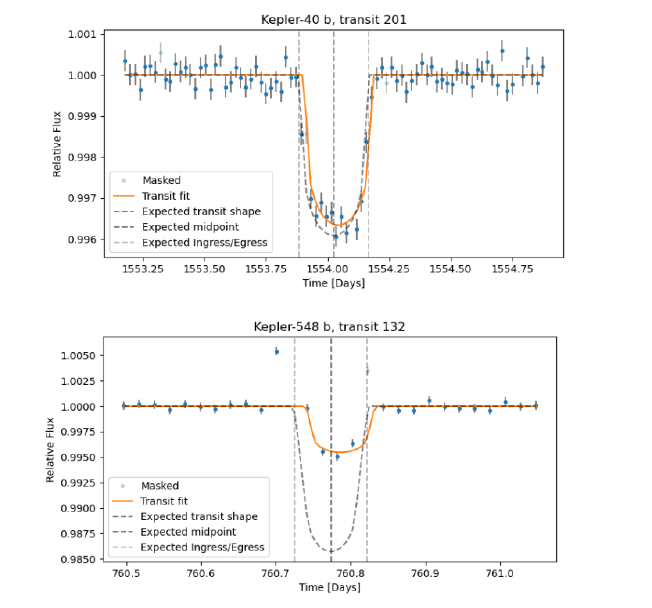}
\caption{Kepler-40b transit 201 and Kepler 548b transit 132 are flagged for large TTVs and pass the visual inspection of the fit. }
\label{fig:ttvs}
\end{figure*}

This leaves two remaining transits, which are shown in Figure \ref{fig:ttvs}.

\begin{figure*}[ht]
\centering
\includegraphics[width=0.75\textwidth]{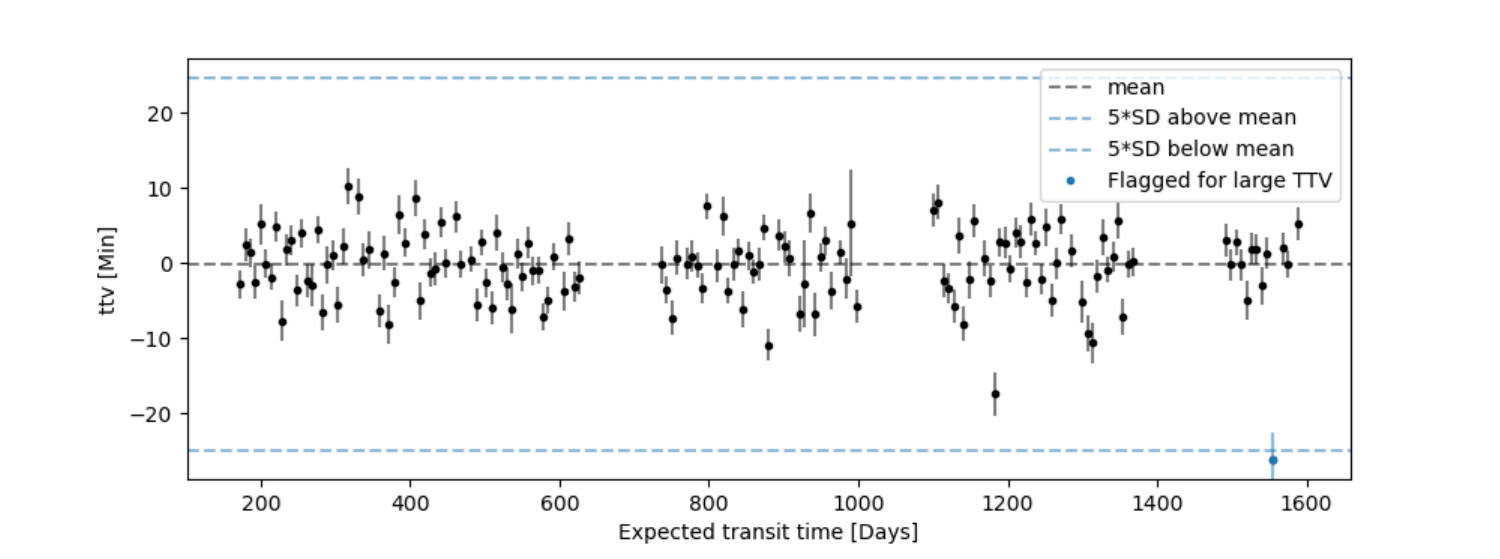}
\caption{Kepler-40b transit 201 is barely above the threshold for flagging, and consistent with contamination from stellar variability}
\label{fig:ttv_flags}
\end{figure*}

Transit 201 is only 5\% above the 5 sigma threshold, but does not obviously fall on the expected tail of a normal distribution of TTVs, as shown in Figure \ref{fig:ttv_flags}. However, Kepler-40b is a rotating variable subgiant star, and thus likely to exhibit starspots. Transit 201 is consistent with the shape expected from contamination due to the exoplanet crossing a starspot feature \citep{Morris2017}.

\begin{figure*}[ht]
\centering
\includegraphics[width=\textwidth]{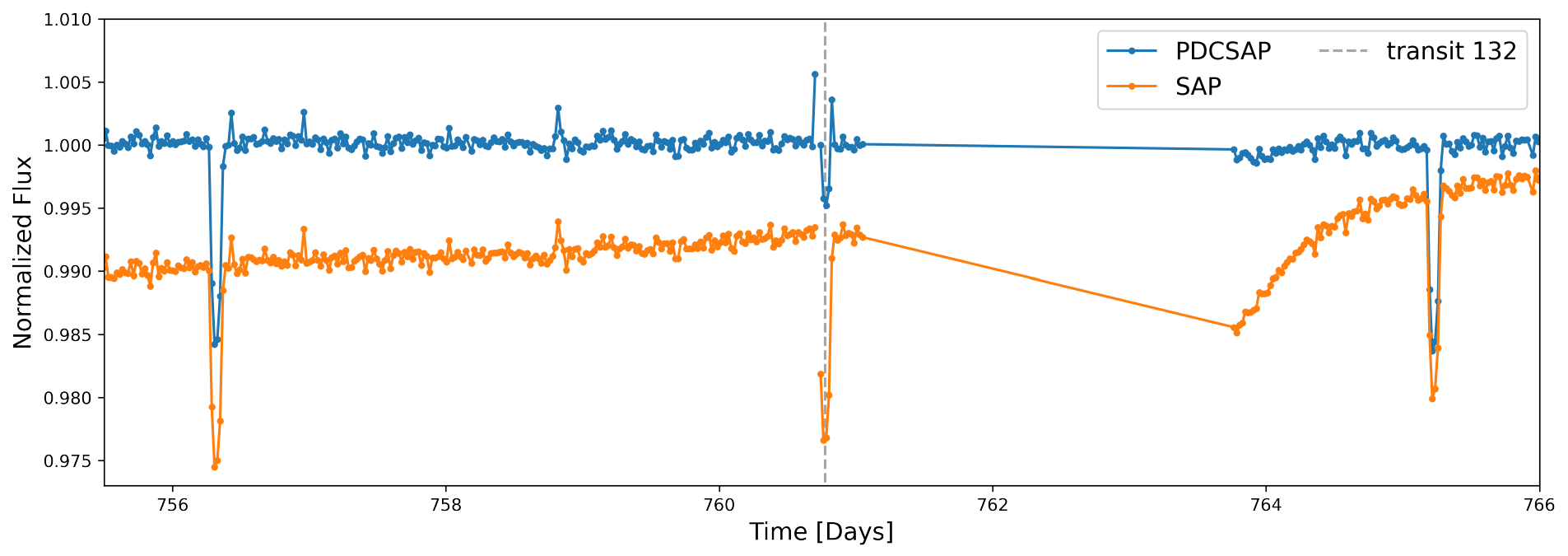}
\caption{Kepler-548b, transit 132, is an apparently very unusual transit with both a large TTV and a shallow depth. However, the transit appears to have a normal shape when viewed using the pre-cotrending SAP flux values, and thus the anomaly can be credited to artifact of the Kepler cotrending process.}
\label{fig:kep548}
\end{figure*}

Kepler-548b transit 132 is flagged for its large TTV, but is most unusual in its dramatically reduced depth. Transits in the system are generally fit well, and no other transit exhibits a shallow depth. No unexpected sibling transit appears in the fitting window (nor could this explain both the large TTV and the shallow depth), and the lightcurve appears free of stellar flaring or other activity which could give the transit an apparently shallower depth. However, when the lightcurve for this transit is created using the raw (SAP) flux values \citep{Twicken2010, Morris2020}, rather than the PDCSAP flux values, the transit no longer appears unusually shallow (Figure \ref{fig:kep548}). This appears to be a rare case in which cotrending applied to the PDCSAP values significantly modifies the shape of a transit, producing a spurious flag. 

\section{Discussion and Conclusions}

To the best of our knowledge, no past work has attempted to characterize the shapes of transits individually in stellar lightcurves, though some have searched individually for TTVs. We have developed and implemented an algorithm to search stellar lightcurves for unusual transits, and have performed a search for such transits on \NSearched\ Kepler transiting exoplanet systems. We find and vet \NMissing\ candidate transits flagged by our algorithm as missing, \NTTV\ flagged as having an unusually large TTV, and \NDeep\ flagged as having an unusually large depth. We visually inspect all flagged transits, and reject each as a likely indication of new astrophysics or extraterrestrial intelligence.  

Our nondetection allows us to put a useful upper limit on the occurrence rate of transits with inexplicably anomalous properties. Zero of \NSearched\ stars host a transit which passed our search algorithm and could not be rejected via manual inspection; the upper limit on detections at 95\%\ confidence with zero detected is 2.99. We conclude that less than 2.99 per \NSearched\ Kepler stars, or less than 1.4\%, host transits with inexplicably large depths or TTVs, or transits which appear to be missing. In choosing our sample we are biased towards planets with deep transits and lightcurves with low flux uncertainties, but this should not bias our detection rates. We are however only sensitive to a small subset of all transits of each planet due to finite Kepler observing times. The median duration of a lightcurve in our target set is 1459.5 days. Thus, our upper limit should be understood to reflect an occurrence rate per 1459.5 days.

Our search is designed to be agnostic to the cause of any anomalous transit shapes. We attempt to reject known astrophysical causes, for instance by rejecting transits where sibling planets, stellar flares, or short-timescale stellar variability can cause apparent anomalies. However, had we been unable to reject any flagged transits an unexpected astrophysical explanation would still be possible. Our search would also be sensitive to non-astrophysical transit anomalies, which represent a possible technosignature in the Search for Extraterrestrial Intelligence. A civilization might choose to cloak its transits from known neighbors, or else broadcast its presence intentionally or unintentionally by modifying its transit shape with megastructures or lasers \citep{Wright2019}. 
While \citet{kipping2016} have noted directed emission from a  laser  could ``cloak'' or modify a transit of a planet in the Kepler band using technology presently available, multi-band detection of transits would be able to distinguish this type of variability compared to megastructures.
In addition, if an intelligent civilization aware of life near our Sun were to attempt to send a beacon or information towards the Sun, it would have to choose the best time to do so. The interval in which the civilization's home planet transits between its host star and the Sun would be clever choice of timing if this civilization expects others to be actively studying the transits of exoplanets for non-SETI purposes. Thus, the interval around a transit is an especially important time to search for SETI signals \citep{Franz2022}. 

\subsection{Future Directions}
The techniques implemented in the work could be readily applied to other large datasets, such as the TESS catalog of transiting exoplanet lightcurves \citep{Borucki2010}. 
In addition, the set of transit fits produced as a result of this work could be very useful as a training dataset in a machine learning search for unexpected transits in stellar lightcurves. Past exoplanet searches miss transits from planets with periods longer than the survey observing time, comets or asteroids, or 
rogue planets. The nature and prevalence of unexpected transits is poorly understood. This represents a major gap in our understanding of the occurrence rate of long-period exoplanets. These long-period planets are likely to be gas giants whose key role in the formation and evolution of planetary systems is an active area of research. Each fit transit from this work can become a labelled positive observation, and fit section of lightcurve with no known transit labelled negative observations, to train a machine learning algorithm to search pixel-by-pixel through stellar lightcurves for transits which appear where none is expected.

Future work could improve on some of the methods used here in order to reduce the need for visual vetting, or to expand the transit fitting approaches. For instance, a more sophisticated method of jointly fitting transits and stellar variability using Gaussian Processes or wavelet analysis could better detrend each transit fitting window and thus mitigate some false positive candidates we encountered. In addition, a future work might benefit from jointly fitting the transits of each sibling planet in a system in order to avoid the necessity of masking the transits of sibling transits in each fitting.


\begin{acknowledgments}
We thank the anonymous referee for helpful comments which improved the final version of this work. 

We thank The Breakthrough Listen Initiative, and everyone who has provided support for this project. 

We are grateful to the Berkeley SETI Research Center (BSRC) for their support of this research. We thank the staff and students of Breakthrough Listen for their support of SETI data collection and making this data publicly available at \url{https://seti.berkeley.edu/opendata}. Breakthrough Listen is managed by the Breakthrough Initiatives, sponsored by the Breakthrough Prize Foundation. 

AZ was funded as a participant in the Berkeley SETI Research Center Research Experience for Undergraduates Site, supported by the National Science Foundation under Grant No.~1950897.

JRAD acknowledges support from the DiRAC Institute in the Department of Astronomy at the University of Washington. The DiRAC Institute is supported through generous gifts from the Charles and Lisa Simonyi Fund for Arts and Sciences, and the Washington Research Foundation.

This paper includes data collected by the Kepler mission. The Kepler mission was a PI-led Discovery Class Mission funded by the NASA Science Mission directorate.

\end{acknowledgments}

%



\software{\texttt{Batman} Transit Model \citep{Kreidberg_2015}, 
\texttt{Lightkurve} \citep{lightkurve},
\texttt{Astropy} \citep{astropy:2013, astropy:2018, astropy:2022}}

\appendix
\section{Summary Table}


\begingroup
\setlength{\tabcolsep}{2pt}
\begin{table}[H]
\centering
\begin{tabular}{ccccccccccc}
\multicolumn{3}{c}{}                                                             & \multicolumn{6}{c}{Median Fit Parameters}                                                         & \multicolumn{2}{c}{Injection/recovery Summary} \\ \hline
\multirow{2}{*}{Name} & \multirow{2}{*}{Siblings} & \multirow{2}{*}{\# Transits} & \multirow{2}{*}{Rp/Rs} & Semimajor     & Inc.      & \multirow{2}{*}{Ecc.} & Arg. of    & Rel.    & Depth Recovery          & TTV recovery         \\
                      &                           &                              &                        & axis {[}AU{]} & {[}Deg{]} &                       & Periastron & Depth   & RMS {[}Days{]}          & RMS                  \\ \hline
Kepler-117 b          & c                         & 77                           & 0.046                  & 21.55         & 89.47     & 0.0                   & 254.3      & 0.00252 & 0.00041                 & 0.313                \\
Kepler-396 c          & b                         & 16                           & 0.049                  & 79.64         & 89.47     & 0.0                   & 0.0        & 0.00245 & 0.00115                 & 0.437                \\
Kepler-782 b          & None                      & 9                            & 0.026                  & 104.61        & 90.0      & 0.0                   & 0.0        & 0.00085 & 0.00014                 & 0.49                 \\
Kepler-39 b           & None                      & 66                           & 0.091                  & 23.72         & 88.42     & 0.0                   & 98.9       & 0.00887 & 0.00139                 & 0.252                \\
Kepler-108 c          & b                         & 7                            & 0.031                  & 70.92         & 89.47     & 0.0                   & 0.0        & 0.00101 & 0.00016                 & 0.595                \\
\end{tabular}
\caption{System information, including median transit fit parameter values, and injection and recovery statistics for each exoplanet in our sample. A subset is shown here as an example. The full table is available in machine readable format.}
\label{tab:summary_table}
\end{table}
\endgroup

\bibliography{works_cited}{}
\bibliographystyle{aasjournal}





\end{document}